\documentclass[a4paper,11pt]{article}
                                                                                                                                                                                                                                                                                                                                                                                                                                                                                                                                                           
\usepackage{amssymb,amsthm,amscd, amsbsy, array,amsmath,mathdots}
\usepackage[francais]{babel}
\usepackage{fontenc}
\usepackage{amsfonts}
\usepackage[mathscr]{eucal}
\usepackage{mathrsfs}
\usepackage{latexsym}
\usepackage{graphicx}
\usepackage[a4paper,textwidth=17cm,textheight=21cm]{geometry}

\newcommand{\be}{\begin{equation}}
\newcommand{\ee}{\end{equation}}
\newcommand{\bea}{\begin{eqnarray}}
\newcommand{\eea}{\end{eqnarray}}

\def\1{{\mbox{\boldmath $1$}}} %
\newcommand{\R}{\mathbb{R}}
\newcommand{\C}{\mathbb{C}}

\newtheorem{theorem}{Theorem}

\begin{document}
\thispagestyle{empty}
\begin{center}
	\begin{huge}
		Noncommutative geometry on the universal envelopping algebra of the Borel subgroup $U(\mathfrak{sb}(2,\C))$
	\end{huge}
\end{center}

\begin{center}
	\begin{large}
		ARM Boris \\
	\end{large}
	{\footnotesize  \textit{arm.boris@gmail.com}}
\end{center}

\vspace{5cm}
\begin{center}
	\textbf{Abstract} \\
	We study the Borel algebra define by $[x_a,x_b]=2\lambda \delta_{a,1} x_b$ \\ as a noncommutative manifold $\R^3_\lambda$. We calculate its \\ noncommutative 
	 differential form relations. We deduce \\ its partial derivative relations and the derivative \\
	  of a plane wave. After calculating its de Rham \\ cohomology,
	   we deduce the wave operator and \\ its corresponding magnetic solution.
\end{center}

\newpage

\setlength{\textheight}{21cm}
\addtolength{\voffset}{0.2cm}

\section{Introduction}
In this paper, we start with the Borel  algebra $\mathfrak{sb}(2,\C)$
\begin{equation}
[J_a,J_b]=2\delta_{a,1} J_b
\end{equation}
If ones believes in Born reciprocity, then there is also theoretical possibility of a sphere, which corresponds to the algebra
\begin{equation}
[x_a, x_b]=2\lambda \delta_{a,1} x_b
\end{equation}
In this paper, we study this algebra as a noncommutative manifold $\R^3_\lambda$. In fact, we consider $x_i$ as coordinates of a noncommutative position space with $\lambda$ the length dimension.
\paragraph{}
In section $2$, we introduce modern quantum group method: we define the Hopf algebras $U(\mathfrak{sb}(2,\C))$ and $\C (SB(2,\C))$ and the pairing between them. We explicit the different action acting on them.
\paragraph{}
In section $3$, we apply this action to the quantum double $D(U(\mathfrak{sb}(2,\C)))=\C(SB(2,\C))_{Ad^*_L}\rtimes U(\mathfrak{sb}(2,\C))$.
\paragraph{}
In section $4$, with a representation of $\mathfrak{sb}(2,\C)$, we can calculate the noncommutative relations between $1$-forms which gives 
\begin{equation}
[x_a,\mathrm{d}x_b]=\delta_{a,1}(1-\delta_{b,1})\lambda \mathrm{d}x_b -\delta_{b,1} \lambda \mathrm{d}x_a 
\label{relationcommforme}
\end{equation}
and the derivative of a general monomial  (\ref{derivativegeneralmonomial}).
With relations (\ref{relationcommforme}), we can deduce the expression of partial derivative (\ref{partialderivative}) and calculate the derivative of wave
\begin{equation}
\mathrm{d} e^{ik.x}=\mathrm{d}x. ik e^{-i\lambda k_1} e^{ik.x}
\label{derivativeplanewavedeux}
\end{equation}
Next, we show that the noncommutative de Rham cohomology of $\R^3_\lambda$ is given by
\begin{equation}
H^0=\C.1, H^1=H^2=H=3=\{ 0 \}
\end{equation}
\paragraph{}
In section $5$, we compute the Hodge $*$-operator. Finally, in section $6$, $7$ and $8$, we compute the wave operator of a plane wave, the kernel of this operator and the gauge potential of a magnetic solution.
\paragraph{}
This paper is widely inspired of the work of \cite{usu} which is done for $U(\mathfrak{su}(2))$ instead of $U(\mathfrak{sb}(2))$

\newpage
\section{Mathematical preliminaries}
\paragraph{}
The Iwasawa decomposition allows us to decompose:
\begin{equation}
\mathfrak{sl}(2,\C)=\mathfrak{su}(2,\C)\oplus \mathfrak{sb}(2,\C)
\label{Iwasawa}
\end{equation}
where $\mathfrak{sb}(2,\C)$ is the Lie algebra of the Borel subgroup $SB(2,\C)$ with commutation relations:
\begin{equation}
[ J_1,J_2]=2 J_2 ,\hspace{2mm}[ J_1,J_3]=2 J_3, \hspace{2mm}[ J_2,J_3]=0
\label{relationcommutation}
\end{equation}
\paragraph{}
The element of $\mathfrak{sb}(2,\C)$ are generated by
 \begin{eqnarray}
 J_1=
\left(\begin{array}{cc}
1&0\\
0&-1\\ 
   \end{array}
\right)
,\hspace{3mm} J_2=
\left(\begin{array}{cc}
0&1\\
0&0\\  
   \end{array}
\right)
,\hspace{3mm}J_3=
\left(\begin{array}{cc}
0&i\\
0&0\\   
   \end{array}
\right)
\label{generateur}
\end{eqnarray}

Here we outline some notions from quantum group theory into which our example fits. For Hopf algebras (i.e. quantum groups), we use the convention of \cite{foundationQGT} . 
It means an algebra $H$ equipped with a coproduct $\Delta :H \to H\otimes H$, counit $\epsilon :H \to \C$ and antipode $S: H \to H$. We will sometimes use the formal  sum notation $\Delta (a) =\sum a_{(1)} \otimes a_{(2)}$, for any $a \in H$. The usual universal enveloping algebra $U(\mathfrak{sb}(2,\C))$ has a structure of cocommutative Hopf algebra generated by $1$ and $J_a, a=1,2,3$ with relations (\ref{relationcommutation}) and 
\begin{equation}
\Delta (J_a) =J_a \otimes 1+1\otimes J_a,\hspace{6mm} \epsilon (J_a)=0,\hspace{6mm} S(J_a)=-J_a
\end{equation}
We also recall that for an Abelian groups, for each Hopf algebra there is a dual one where the product of one is adjoint to the coproduct of the other. $U(\mathfrak{sb}(2,\C))$ is dually  paired with the commutative Hopf algebra $\C(SB(2,\C))$ generated by coordinate functions $t^i_{\hspace{2mm}j}$, for $i,j=1,2$ on $SB(2,\C)$ satisfying the determinant relation $t^1_{\hspace{2mm}1} t_{\hspace{2mm}2}^2 -t^1_{\hspace{2mm}2} t^2_{\hspace{2mm}1} =1$ and with:
\begin{equation}
\Delta (t^i_{\hspace{2mm}j})= \sum_{k=1}^2 t^i_{\hspace{2mm}k} \otimes t^k_{\hspace{2mm}j}, \hspace{4mm}\epsilon(t^i_{\hspace{2mm}j})=\delta_j^i,\hspace{4mm} St^i_{\hspace{2mm}j}=(t^i_{\hspace{2mm}j})^{-1}
\end{equation}
where inversion is an algebra-valued matrix. The pairing between the algebras $U(\mathfrak{sb}(2,\C))$ and $\C(SB(2))$ is defined by
\begin{equation}
<\xi, f >=\frac{d}{dt} f(e^{t\xi})|_{t=0}
\end{equation}
where $\xi \in \mathfrak{sb}(2,\C)$ and $f \in \C (SB(2,\C))$ which results in particular in:
\begin{equation}
<J_a, t^i_{\hspace{2mm}j}>=J^{\hspace{2mm} i}_{a \hspace{2mm}j}
\end{equation}
where $J^{\hspace{2mm} i}_{a \hspace{2mm}j}$ are the $i,j$ entries of the matrix $J_a, a=1..3$.
\paragraph{}
We also need standard notions of actions and coactions. A left coaction of a Hopf algebra $H$ on a space $V$ means a map $V \to H\otimes V$ obeying axioms like those of an action but reversing all maps. So a coaction of $\C (SB(2,\C))$ essentially corresponds to an action of $U(\mathfrak{sb}(2,\C))$ via the pairing. Examples are:
\begin{equation}
Ad_L (h)(g)=h\rhd g= \sum h_{(1)} g S(h_{(2)})
\label{leftadjointaction}
\end{equation}
the left adjoint action.
$Ad_L :H\otimes H \to H$. Its arrow-reversal is the left adjoint coaction $Ad^L : H \to H\otimes H$,
\begin{equation}
Ad^L (h)(g)= \sum h_{(1)}  S(h_{(3)})\otimes h_{(2)}
\end{equation}
There are also the regular action (given by the product), regular coaction (given by $\Delta : H \to H\otimes H$), and coadjoint actions and coregular actions of the dual, given via the pairing from the adjoint and regular coactions, etc \cite{foundationQGT}.
We will nedd the left coadjoint action of $H$ on a dual quantum group $A$:
\begin{equation}
Ad_L^*(h)(\phi)=h\rhd \phi =\sum \phi_{(2)} <(S\phi_{(1)}) \phi_{(2)}, h>,\hspace{3mm} \forall h \in H,\hspace{3mm} \phi \in A
\label{leftcoadjointaction}
\end{equation}
and the right coregular action of $A$ on $H$ which we will view as a left action of the opposite algebra $A^{op}$:
\begin{equation}
\phi \rhd h =\sum <\phi, h_{(1)}> h_{(2)}, \hspace{2mm} \forall h \in H ,\hspace{2mm} \phi \in A
\label{coregular}
\end{equation}
Given a quantum group $H$ dual to a quantum group $A$, there is a quantum double written loosely as $D(H)$ and containing $H, A$ as sub-Hopf algebras. More precisely it is a double cross product $A^{op} \bowtie H$ where there are cross relations given by a mutual coadjoint actions \cite{foundationQGT}. Also, $D(H)$ is formally quasitriangular in the sense of a formal 'universal $R$ matrix' $\mathcal{R}$ with terms in $D(H) \otimes D(H)$. The detailed structure of $D(U(\mathfrak{sb}(2,\C)))$ is covered  in Section $3$ and in this case is more simply a semidirect product $\C (SB(2,\C))\rtimes U(\mathfrak{sb}(2,\C))$  by the coadjoint action. 
\paragraph{}
Finally, we will need  the notion of differential calculus on an algebra $H$. This is common to several approaches to noncommutative geometry including that of Connes \cite{Connes} . A first order calculus means to specify $(\Omega^1,\mathrm{d})$, where $\Omega^1$ is an $H-H$-bimodule, $\mathrm{d}:H\to \Omega^1$ obeys the Leibniz rule,
\begin{equation}
\mathrm{d}(hg)=(\mathrm{d}h)g+h(\mathrm{d}g)
\end{equation}
and $\Omega^1$ is spanned by elements of the form $(\mathrm{d}h)g$. A bimodule just means that one can multiply '$1$-forms' in $\Omega^1$ by 'functions' in $H$ from the left and right coactions of $H$ in $\Omega^1$ (a bicomodule) which are themselves bimodule homomorphisms, and d interwines the coactions with the regular coactions of $H$ on itself. Given a bicovariant calculus one can find invariant forms
\begin{equation}
\omega (h)=\sum (\mathrm{d}h_{(1)}) Sh_{(2)}
\label{omegah}
\end{equation}
for any $h \in H$. The span of such invariant forms is a space $\Lambda^1$ and all of $\Omega^1$ can be reconstructed from them via 
\begin{equation}
\mathrm{d}h=\sum \omega (h_{(1)})h_{(2)}
\end{equation}
As a result, the construction of a differential structure on a quantum group rests on that of $\Lambda^1$, with $\Omega^1=\Lambda^1.H$. They 
in turn can be constructed in the form
\begin{equation}
\Lambda^1 =\ker \epsilon /\mathcal{I}
\end{equation}
where $\mathcal{I} \subset \ker \epsilon$ is some left  ideal in $H$ that is $Ad^L$-stable \cite{Woronwicz}. We will use this method in Section $4$ to introduce  a reasonable  calculus  on $U(\mathfrak{sb}(2,\C))$. Some general remarks (but not our calculus, which seems to be new) appared in \cite{riemanngeometry}.
\paragraph{}
Any bicovariant calculus has a 'minimal' extension to an entire exterior algebra \cite{Woronwicz}. One uses the universal R-matrix of the quantum double to define a braiding operator on $\Lambda^1 \otimes \Lambda^1$ and uses it to 'antisymmetrize' the formal algebra generated by the invariant forms. These and elements of $H$ define $\Omega$ in each degree. In our case of $U(\mathfrak{sb}(2,\C))$, because it is cocommutative, the braiding is the usual flip. Hence we have the usual anticommutation relations among invariant forms. We also extend d:$\Omega^k \to \Omega^{k+1}$ as a (right-handed) super derivation by:
\begin{equation}
\mathrm{d}(\omega \wedge \eta )=\omega \wedge \mathrm{d}\eta +(-1)^{\mathrm{deg}\eta} \mathrm{d} \omega \wedge \eta
\end{equation}
A differential calculus is said to be inner if the exterior differentiation in $\Omega^1$ ( and hence in all degrees) is given by the (graded) commutator with an invariant $1$-form $\theta \in \Lambda^1$, that is
\begin{equation}
\mathrm{d}\omega =\omega \wedge \theta -(-1)^{\mathrm{deg}\omega} \theta \wedge \omega
\end{equation}
Almost all noncommutative geometries that one encounters are inner, which is the fundamental reason that they are in many ways better behaved than the classical case.

\section{The Qantum Double as Exact Isometries of $\R^3_\lambda$}
In this section we first of all recall the structure of the quantum double $D(U(\mathfrak{sb}(2,\C)))$ in the context of Hopf algebra theory. We will then explain its canonical action on a second copy $\R^3_\lambda \cong U(\mathfrak{sb}(2,\C))$ arising from the general Hopf algebra theory, thereby presenting it explicitly as an exact quantum symmetry group of that. Here $x_a =\lambda J_a$ is the isomorphism valid  for $\lambda \neq 0$. By an exact quantum symmetry we mean that the quantum group acts on $\R^3_\lambda$ with the product of $\R^3_\lambda$  an intertwiner (i.e. the algebra is covariant).
Because $U(\mathfrak{sb}(2,\C))$  is cocommutative, its quantum double $D(U(\mathfrak{sb}(2,\C)))$ is a usual crossed product \cite{foundationQGT}
\begin{equation}
D(U(\mathfrak{sb}(2,\C)))=\C(SB(2,\C))_{Ad_L^*}\rtimes U(\mathfrak{sb}(2,\C))
\end{equation}
where the action  is induced by the adjoint action (it is the coadjoint action on $\C(SB(2,\C))$). This crossed product is isomorphic as a vector space with $\C(SB(2,\C))\otimes U(\mathfrak{sb}(2,\C))$ but with algebra  structure given by
\begin{equation}
(a\otimes h)(b\otimes g)=\sum a Ad^*_{_L h_{(1)}} (b) \otimes h_{(2)}g
\end{equation}
for $a,b \in \C(SB(2,\C))$ and $h,g \in U(\mathfrak{sb}(2,\C))$. In terms of the generators, the left coadjoint action (\ref{leftcoadjointaction}) takes the form 
\begin{equation}
Ad_{_L J_a} (t^i_{\hspace{2mm}j})=\sum t^k_{\hspace{2mm}l} <J_a, S(t^i_{\hspace{2mm}k})t^l_{\hspace{2mm}j}>=t^i_{\hspace{2mm}k} J_{a \hspace{3mm}l}^{\hspace{2mm}k}-J_{a \hspace{3mm}k}^{\hspace{2mm}i}t^k_{\hspace{2mm}j}
\end{equation}
As a result we find that $D(U(\mathfrak{sb}(2,\C))$  is generated by $U(\mathfrak{sb}(2,\C))$ and $\C(SB(2,\C))$ with cross relations 
\begin{equation}
[J_a, t^i_{\hspace{2mm}j}]=t^i_{\hspace{2mm}k} J_{a \hspace{3mm}l}^{\hspace{2mm}k}-J_{a \hspace{3mm}k}^{\hspace{2mm}i}t^k_{\hspace{2mm}j}
\end{equation}
Meanwhile the coproducts are the same as those of $U(\mathfrak{sb}(2,\C))$ and $\C (SB(2,\C))$.
Next, a general feature of any quantum double is a canonical or 
'Schrodinger' representation, where $U(\mathfrak{sb}(2,\C))\subset 
D(U(\mathfrak{sb}(2,\C)))$ acts on $U(\mathfrak{sb}(2,\C))$ by the left 
adjoint action (\ref{leftadjointaction}) and $\C (SB(2,\C)) \subset 
D(U(\mathfrak{sb}(2,\C)))$ acts by the coregular one (\ref{coregular}), see 
\cite{foundationQGT}. We denote the acted-upon copy by $\R^3_\lambda$. Then 
$J_a$ simply acts by 
\begin{equation}
J_a \rhd f(x) =\lambda^{-1} \sum x_{a(1)} f(x) S(x_{a(2)})=\lambda^{-1}[x_a,h],\forall f(x) \in \R^3_\lambda
\end{equation} 
e.g.
\begin{equation}
J_a\rhd x_b=2\delta_{a,1} x_b 
\end{equation}
with the co-regular  action reads
\begin{equation}
t^i_{\hspace{2mm}j}\rhd f(x)=<t^i_{\hspace{2mm}j},f(x)_{(1)}>f(x)_{(2)},\hspace{6mm} e.g. \hspace{6mm}  t^i_{\hspace{2mm}j}\rhd x_a =\lambda  J_{a \hspace{3mm}k}^{\hspace{2mm}i} 1+ \delta^i_{\hspace{2mm}j} x_a
\end{equation}
The general expression is given by a shuffle product (see Section $4$). \textit{With this action, $\R^3_\lambda$ turns into a left $D(U(\mathfrak{sb}(2,\C)))$-covariant algebra.}

\section{The 3-Dimensional Calculus on $\R^3_\lambda$}
The purpose of this section is to construct a bicovariant calculus on the algebra $\R^3_\lambda$ following the steps outlined in Section $2$, the calculus we obtain being that on the algebra $U(\mathfrak{sb}(2,\C))$ on setting $\lambda =1$. We write $\R^3_\lambda$ as generated by $x_1, x_2$ and $x_3$, say, and with the Hopf algebra structure given explicitly in terms of the generators as
\begin{equation}
[x_1,x_2]=2\lambda x_2; \hspace{3mm} [x_1,x_3]=2\lambda x_3; \hspace{3mm} [x_2,x_3]=0 
\label{relationdecommutation}
\end{equation}
and the additive coproduct  as before. The particular form of the coproduct, the relations and  (\ref{omegah}) shows that d$\xi=\omega(\xi)$ for all $\xi \in \mathfrak{sb}(2,\C)$. Because  of the cocommutativity, all ideals in $\R^3_\lambda$ are classified simply by the ideals $\mathcal{I} \subset \ker \epsilon$. In general the coirreductible calculi (i.e. having no proper quotients) are labelled by pairs $(V_\rho,\Lambda)$, with $\rho:U(\mathfrak{g}) \to$End$V_\rho$ an irreductible representation of $U(\mathfrak{g})$ and $\Lambda$ a ray in $V_\rho$. In order to construct an ideal of $\ker \epsilon$, take the map
\begin{equation}
\rho_\Lambda : U(\mathfrak{g}) \to V_\rho, \hspace{3mm} h \longmapsto \rho(h). \Lambda
\end{equation}
It is easy to see that ker$\rho_\Lambda$ is a left ideal in ker $\epsilon$. Then, if $\rho_\Lambda$ is surjective, the space of $1$-forms can be identified with $V_\rho =\ker \epsilon / \ker \rho_\Lambda$. The general commutation relations are 
\begin{equation}
av=va+\rho (a) .v
\label{generalrelationcommutation}
\end{equation}
and the derivative for a general monomial $\xi_1 ... \xi_n$ is given by the expression
\begin{equation}
\mathrm{d}(\xi_1 ... \xi_n)=\frac{1}{\lambda} \sum_{k=1}^n \sum_{\sigma \in S(n,k)} \rho_\Lambda (\xi_{\sigma (1)}...\xi_{\sigma(k)})\xi_{\sigma(k+1)}...\xi_{\sigma(n)}
\end{equation}
We explore some examples of coirreductible calculi for the universal enveloping algebra $\R^3_\lambda$, generated by $x_1, x_2$ and $x_3$ satisfying (\ref{relationdecommutation}). First, let us analyse the three dimensional, coirreductible calculus on $\R^3_\lambda$ by taking 
 $V_\rho =\C^3$, with basis
 \begin{eqnarray}
e_1 =
\left(\begin{array}{c}
1\\
0\\ 
0\\  
   \end{array}
\right)
,\hspace{3mm} e_2 =
\left(\begin{array}{c}
0\\
1\\ 
0\\  
   \end{array}
\right)
,\hspace{3mm} e_3 =
\left(\begin{array}{c}
0\\
0\\ 
1\\  
   \end{array}
\right)
\label{base}
\end{eqnarray}
In this basis, the representation $\rho$ takes the form
 \begin{eqnarray}
 \rho(x_1)=\lambda
\left(\begin{array}{ccc}
1&0&0\\
0&1&0\\ 
0&0&-1\\  
   \end{array}
\right)
,\hspace{3mm} \rho(x_2)=\lambda
\left(\begin{array}{ccc}
0&0&1\\
0&0&0\\ 
0&0&0\\  
   \end{array}
\right)
,\hspace{3mm}\rho (x_3)=\lambda
\left(\begin{array}{ccc}
0&0&0\\
0&0&1\\ 
0&0&0\\  
   \end{array}
\right)
\label{representation}
\end{eqnarray}
We choose, for example, $\Lambda=e_3$. The space of $1$-forms will be generated by the vectors $e_1, e_2$ and $e_3$. The derivative of the generators of the algebra are given by 
\begin{equation}
\mathrm{d}x_1=\lambda^{-1} \rho(x_1).e_3=-e_3,\hspace{3mm} \mathrm{d}x_2=\lambda^{-1} \rho(x_2).e_3=e_1,\hspace{3mm}
\mathrm{d}x_3=\lambda^{-1} \rho(x_3).e_3=e_2
\end{equation}
The commutation relations between the basic $1$-forms and the generators can be deduced from (\ref{generalrelationcommutation}) giving 
\begin{eqnarray}
\nonumber x_1 e_1=e_1 x_1+\lambda e_1 \\
\nonumber x_1 e_2=e_2 x_1 +\lambda e_2 \\
\nonumber x_1 e_3=e_3 x_1-\lambda e_3\\
\nonumber x_2 e_1=e_1 x_2\\
\nonumber x_2 e_2=e_2 x_2\\
\nonumber x_2 e_3 =e_3 x_2 +\lambda e_1\\
\nonumber x_3 e_1 =e_1 x_3\\
\nonumber x_3 e_2 =e_2 x_3\\
 x_3 e_3=e_3x_3+\lambda e_2
\end{eqnarray}
The compability conditions of this definition of the derivative with the Leibniz rule is due to the following commutation relations between the generators of the algebra and the basic $1$-forms:
\begin{eqnarray}
\nonumber x_1 \mathrm{d}x_2=\mathrm{d}x_2 x_1 +\mathrm{d}x_2\\
\nonumber x_1 \mathrm{d}x_3=\mathrm{d}x_3 x_1 +\lambda \mathrm{d}x_3 \\
\nonumber x_1 \mathrm{d}x_1=\mathrm{d}x_1 x_1-\lambda \mathrm{d}x_1\\
\nonumber x_2 \mathrm{d}x_2=\mathrm{d}x_2 x_2\\
\nonumber x_2 \mathrm{d}x_3=\mathrm{d}x_3 x_2\\
\nonumber x_2 \mathrm{d}x_1 =\mathrm{d}x_1 x_2 -\lambda \mathrm{d}x_2\\
\nonumber x_3 \mathrm{d}x_2 =\mathrm{d}x_2 x_3\\
\nonumber x_3 \mathrm{d}x_3 =\mathrm{d}x_3 x_3\\
 x_3 \mathrm{d}x_1=\mathrm{d}x_1 x_3-\lambda \mathrm{d}x_3
 \label{relationcommutationform}
\end{eqnarray}
The commutation relations (\ref{relationcommutationform}) have a simple expression:
\begin{equation}
[x_a, \mathrm{d}x_b]=\delta_{a,1}(1-\delta_{b,1})\lambda \mathrm{d}x_b-\delta_{b,1} \lambda \mathrm{d}x_a
\end{equation}
In the classical limit, this calculus turns out to be the commutative calculus on usual three dimensional Euclidean space. 
In this basis the partial derivatives defined by
\begin{equation}
\mathrm{d}f(x)=(\mathrm{d}x_a)\partial^a f(x)
\label{partial}
\end{equation}
The explicit expression for the derivative of a general monomial $x_1^a x_2^b x_3^c$ is given
\begin{eqnarray}
\nonumber \mathrm{d}(x_1^a x_2^b x_3^c)=-(1-\delta_{a,0})&\mathrm{d}x_1& \sum_{k=1}^a A^k_a (-1)^k \lambda^{k-1} x_1^{a-k}  x_2^b x_3^c\\
\nonumber + &\mathrm{d}x_2& \sum_{k=0}^a A^k_a b \lambda^k x_1^{a-k} x_2^{b-1} x_3^c\\
+&\mathrm{d}x_3& \sum_{k=0}^a A^k_a c \lambda^k x_1^{a-k} x_2^{b} x_3^{c-1}
\label{derivativegeneralmonomial}
\end{eqnarray}
here $A^k_n=\frac{n!}{(n-k)!}$ is the number of $k$ arrangment among $n$. The noncommutative partial derivatives $\partial_a$ defined in (\ref{partial}) have the expressions to lowest order
\begin{eqnarray}
\nonumber \partial_1 f(x)&=&\overline{\partial}_1 f(x)-\lambda \overline{\partial}_1^{\hspace{2mm}2} f(x)\\
\nonumber \partial_2 f(x)&=&\overline{\partial}_2 f(x)-\lambda \overline{\partial}_1  \overline{\partial}_2 f(x)\\
 \partial_3 f(x)&=&\overline{\partial}_2 f(x)-\lambda \overline{\partial}_1  \overline{\partial}_3 f(x)
 \label{partialderivative}
\end{eqnarray}
where $\overline{\partial}_a$ are the usual derivatives in classical variables and we do not write the normal ordering on expressions already $O(\lambda)$ since the error is higher order
\paragraph{}
Next, the expression for the derivatives of plane waves is very simple.    In terms of generators $x_a$, the derivative of the plane wave $e^{i\sum_a k^a x_a}=e^{ik.x}$ is given by
\begin{equation}
\mathrm{d}e^{ik.x}=\mathrm{d}x. ik e^{-i\lambda k_1} e^{ik.x}
\label{derivativeplanewave}
\end{equation}
One can see that the limit $\lambda \to 0$ gives the correct  formula   for the derivative of plane waves, that is 
\begin{equation}
\lim_{\lambda \to 0}\mathrm{d}e^{ik.x}=(\sum_{a=1}^3 ik_a \mathrm{d}x_a)e^{ik.x}=ik.(\mathrm{d}x)e^{ik.x}
\end{equation}
where at $\lambda =0$ on the right hand side we have the classical coordinates and the classical $1$-forms in usual three dimensional commutative calculus. 
We can check that the Casimir operator:
\begin{equation}
C=x_1^2
\end{equation}
have the derivative:
\begin{equation}
\mathrm{d}C=2 \mathrm{d}x_1 (x_1-\lambda)
\end{equation}
\paragraph{}
We can also construct the full exterior algebra 
$\Omega(\R^3_\lambda)=\oplus_{n=0}^\infty \Omega^n(\R^3_\lambda)$. In our 
case the general building \cite{Woronwicz} becomes the trivial flip 
homomorphism because the right invariant basic $1$-forms are also left 
invariant. Hence our basic $1$-forms in $M_2(\C)$ are totally 
anticommutative and their usual antisymmetric wedge product generates the 
usual exterior algebra on the vector space $M_2(\C)$. The full 
$\Omega(\R^3_\lambda)$ is generated by these and  elements of 
$\R^3_\lambda$ with the relations (\ref{relationcommutationform}). 
The cohomologies of this calculus were also calculated giving the following results:
\begin{theorem}
The noncommutative de Rham cohomology of $\R^3_\lambda$ is 
\begin{equation}
H^0=\C.1,\hspace{3mm}H^1=H^2=H^3=\{ 0 \}
\end{equation}
\end{theorem}
\textit{Proof}
This is direct (and rather long) computation of the closed forms and the exact ones in each degree using the explicit formula (\ref{derivativegeneralmonomial}) on general monomials. To give an example of the procedure, we will doit in some detail for the case of $1$-forms. Take a general $1$-form
\begin{equation}
\omega =\alpha (\mathrm{d}x_1)x_1^a x_2^b w_3^c+\beta (\mathrm{d}x_2)x_1^d x_2^e x_3^f+\gamma (\mathrm{d}x_3)x_1^g x_2^h x_3^i
\end{equation}
and impose d$\omega =0$. We start analysing the simplest cases, and then going to more complex ones.
\paragraph{}
Taking $\beta =\gamma  =0$, then 
\begin{equation}
\omega =\alpha (dx_1)x_1^a x_2^b w_3^c
\end{equation}
The vanishing of the term   $\mathrm{d}x_2 \wedge \mathrm{d}x_1$ leads to the conclusion  that $b=0$. Similarly, the vanishing of the term in $\mathrm{d}x_3 \wedge \mathrm{d}x_1$  
leads to $c=0$ so that:
\begin{equation}
\omega =\alpha \mathrm{d}x_1 x_1^a =\frac{\alpha}{a+1} \mathrm{d}(x^{a+1})
\end{equation}
which is an exact form, hence belonging to the null cohomology class.
 The cases $\alpha =\gamma =0$ and $\alpha = \beta =0$ also lead to exact forms.
 \paragraph{}
 Let us now analyse the case with two non zero terms:
 \begin{equation}
\omega =\alpha (\mathrm{d}x_1)x_1^a x_2^b w_3^c+\beta (\mathrm{d}x_2)x_1^d x_2^e x_3^f
\end{equation}
The vanishing condition in the term on $dx_1 \wedge dx_2$ reads
\begin{equation}
-\alpha \sum_{k=0}^a A^k_a b \lambda ^k x_1^{a-k} x_2^{b-1} x_3^c=(1-\delta_{d,0})\beta \sum_{k=1}^d A^k_d (-1)^k \lambda ^{k-1} x_1^{d-k} x_2^e x_3^f
\end{equation}
Then we conclude that $a=0, d=1, \alpha b =\beta, b-1= e, c=f$. The vanishing of the terms $\mathrm{d}x_1 \wedge \mathrm{d}x_3$ and $\mathrm{d}x_2 \wedge \mathrm{d}x_3$
reads $\alpha c =\beta f=0$. So we have $c=f=0$ then 
\begin{eqnarray}
\nonumber \omega &=&\alpha \mathrm{d}x_1 x_2^b+\alpha b \mathrm{d}x_2 x_1 x_2^{b-1}\\
\nonumber &=&\alpha[\mathrm{d}(x_1 x_2^b)-\mathrm{d}x_2 b \lambda x_2^{b-1}]\\
\nonumber \omega &=&\alpha \mathrm{d}(x_1 x_2^b -\lambda x_2^b)
\end{eqnarray}
which is exact.
\paragraph{}
Now consider the case 
\begin{equation}
\omega =\beta (\mathrm{d}x_2)x_1^d x_2^e x_3^f +\gamma (\mathrm{d}x_3)x_1^g x_2^h x_3^i
\end{equation}
The vanishing condition in the term on $\mathrm{d}x_2 \wedge \mathrm{d}x_3$ reads
\begin{equation}
\beta \sum_{k=0}^d A^k_d f \lambda ^k x_1^{\lambda -k} x_2^e x_3^{f-1}=\gamma \sum_{k=0}^g A^k_g h \lambda ^k x_1^{g-k} x_2^{h-1} x_3^i
\end{equation}
Then we conclude that $\beta f=\gamma h, g=d, h-1=e, f-1=i$. The vanishing of $\mathrm{d}x_1 \wedge \mathrm{d}x_3$ and $\mathrm{d}x_1 \wedge \mathrm{d}x_2$ reads $\gamma g=\beta d=0$. So we have $d=0$ then
\begin{eqnarray}
\nonumber \omega &=&\frac{\beta}{h}(h \mathrm{d}x_2 x_2^{h-1} x_3^f +f \mathrm{d}x_3 x_2^h x_3^{f-1})\\
\nonumber \omega &=&\frac{\beta}{h} \mathrm{d}(x_2^h x_3^f)
\end{eqnarray}
which is exact.
\paragraph{}
Now we consider the $2$-forms:
\begin{equation}
 \omega =\alpha \mathrm{d}x_1 \wedge \mathrm{d}x_2 x_1^a x_2^b x_3^c
\end{equation}
The vanishing condition in the term on $\mathrm{d}x_1 \wedge \mathrm{d}x_2 \wedge \mathrm{d}x_3$ reads $\alpha c=0$. So the form becomes:
\begin{eqnarray}
\omega &=&\alpha \mathrm{d}x_1 \wedge \mathrm{d}x_2 x_1^a x_2^b \\
\omega &=& \alpha \mathrm{d}(\mathrm{d}x_2 \frac{x_1^{a+1}}{a+1}x_2^b)
\end{eqnarray}
which is exact.
\paragraph{}
Now consider the form
\begin{equation}
\omega =\alpha \mathrm{d}x_1 \wedge \mathrm{d}x_2 x_1^a x_2^b x_3^c +\beta \mathrm{d}x_2 \wedge \mathrm{d}x_3 x_1^d x_2^e x_3^f
\end{equation}
The vanishing condition in the term on $\mathrm{d}x_1 \wedge \mathrm{d}x_2 \wedge dx_3$ reads
\begin{equation}
\alpha \sum_{k=0}^a A^k_a c \lambda ^k x_1^{a-k} x_2^b x_3^{c-1}=-\beta \sum_{k=1}^d A^k_d (-1)^k \lambda^{k-1} x_1^{d-k} x_2^e x_3^f
\end{equation}
Then we conclude that $\alpha c =-\beta, d=1, a=0, b=e, c-1=f$. So
\begin{eqnarray}
\omega &=& \alpha \mathrm{d}x_1 \wedge \mathrm{d}x_2 x_2^b x_3^c -\alpha c \mathrm{d}x_2 \wedge \mathrm{d}x_3 x_1 x_2^b x_3^{c-1} \\
\omega &=& \alpha \mathrm{d}([x_1x_2^b x_3^c -\lambda x_2^b x_3^c]\mathrm{d}x_2)
\end{eqnarray}
which is exact.
The proof that all higher cohomologies are trivial is also an exhaustive analysis of all the possible cases and inductions on powers of  $h$, as exemplified here for the $3$-forms. It is clear that all $3$-forms 
\begin{equation}
\omega =\alpha \mathrm{d}x_1 \wedge \mathrm{d}x_2 \wedge \mathrm{d}x_3 x_1^a x_2^b x_3^c
\end{equation}
are closed. We use induction on $n$ to prove that there exists a three form $\eta$ such that $\omega=d\eta$.
\paragraph{}
For $n=0$, we have 
\begin{equation}
\mathrm{d}x_1 \wedge \mathrm{d}x_2 \wedge \mathrm{d}x_3  x_2^b x_3^c=d(\mathrm{d}x_1 \wedge \mathrm{d}x_2 \frac{x_2^b x_3^{c+1}}{c+1})
\end{equation}
Suppose that there exist $3$-forms $\eta_k$, for $0\le k<n$, such that
\begin{equation}
\mathrm{d}x_1 \wedge \mathrm{d}x_2 \wedge \mathrm{d}x_3 x_1^a x_2^b x_3^c=d\eta _a
\end{equation} 
then 
\begin{eqnarray}
\nonumber \mathrm{d}x_1 \wedge \mathrm{d}x_2 \wedge \mathrm{d}x_3 x_1^a x_2^b x_3^c&=& \mathrm{d}x_1 \wedge \mathrm{d}x_2 \wedge \mathrm{d}x_3 [\sum_{k=1}^{a+1} A_{a+1}^k (-1)^k \lambda ^{k-1} x_1^{a+1-k} x_2^b x_3^c \\
\nonumber &&-\sum_{k=2}^{a+1} A_{a+1}^k (-1)^k \lambda ^{k-1} x_1^{a+1-k} x_2^b x_3^c]\\
&=&\mathrm{d}(-\mathrm{d}x_2 \wedge \mathrm{d}x_3 \frac{x_1^{a+1}}{a+1} x_2^b x_3^c -\sum_{k=2}^{a+1} A^k_{a+1} (-1)^k \lambda ^{k-1} \eta_{a+1-k})
\end{eqnarray}
Hence all $3$-forms are exact. The same procedure is used to show the triviality of the other cohomologies.\\
\begin{flushright}
$\spadesuit$
\end{flushright}
\paragraph{}
For $\R^3_\lambda$ we should expect the cohomology to be trivial, since this corresponds to Stokes theorem and many other aspects taken for granted in physics.

\section{Hodge $*$-Operator and Electromagnetic Theory}
The above geometry also admits a metric structure. It is known that any nondegenerate bilinear form $\eta \in \Lambda ^1 \otimes \Lambda ^1$ defines an invariant metric on the Hopf algebra $H$ \cite{riemanngeometry}. For the case of $\R^3_\lambda$ we can define the metric
\begin{equation}
\eta = \mathrm{d}x_1 \otimes \mathrm{d}x_1 +\mathrm{d}x_2 \otimes \mathrm{d}x_2 +\mathrm{d}x_3 \otimes \mathrm{d}x_3
\label{metric}
\end{equation}
for a parameter $\mu$. This bilinear form is non-degenerate, invariant by left and right coactions and symmetric in the sense that $\wedge (\eta)=0$. With this metric structure, it is possible to define a Hodge $*$-operator and then explore the properties of the Laplacian and find some physical consequences. Our picture is similar to \cite{noncommutativecohomology} where the manifold is similarly three dimensional.
\paragraph{}
The Hodge $*$-operator on a $n$-dimensional calculus (for which the top form is of order $n$), over a Hopf algebra $H$ with the metric $\eta$ is 	a map $*:\Omega^k \to \Omega^{n-k}$ given by the expression
\begin{equation}
*(\omega_{i_1}...\omega_{i_k})=\frac{1}{(n-k)!} \epsilon _{i_1...i_k i_{k+1}...i_n} \eta^{i_{k+1}j_1}...\eta^{i_nj_{n-k}}\omega_{j_1}...\omega_{j_{n-k}}
\end{equation}
In the case of the algebra $\R^3_\lambda$, we have a four dimensional calculus with $\omega_1=\mathrm{d}x_1, \omega_2=\mathrm{d}x_2, \omega_3=\mathrm{d}x_3$. The components of the metric inverse, as we can see from (\ref{metric}), are $\eta^{11}=\eta^{22}=\eta^{33}=1$. 
The expressions for the Hodge $*$-operator are summarize as follows:
\begin{eqnarray}
\nonumber *1&=&\mathrm{d}x_1 \wedge \mathrm{d}x_2 \wedge \mathrm{d}x_3 \\
\nonumber *\mathrm{d}x_1&=&\mathrm{d}x_2 \wedge \mathrm{d}x_3\\
\nonumber *\mathrm{d}x_2&=&\mathrm{d}x_3 \wedge \mathrm{d}x_1 \\
\nonumber *\mathrm{d}x_3&=&\mathrm{d}x_1 \wedge \mathrm{d}x_2\\
\nonumber *(\mathrm{d}x_1 \wedge \mathrm{d}x_2)&=&\mathrm{d}x_3\\
\nonumber *(\mathrm{d}x_1 \wedge \mathrm{d}x_3)&=&-\mathrm{d}x_2\\
\nonumber *(\mathrm{d}x_2 \wedge \mathrm{d}x_3)&=&\mathrm{d}x_1\\
 *(\mathrm{d}x_1 \wedge \mathrm{d}x_2 \wedge \mathrm{d}x_3)&=&1
\end{eqnarray}
We note that $**(\omega)=\omega$.
\paragraph{}
Given the Hodge $*$-operator, one can write, for example, the coderivative $\delta =*$d$*$ and the Laplacian operator $\Delta =\delta$d$+$d$\delta$. Note that the Laplacian maps to forms of the same degree. We prefer to work actually with the 'Maxwell-type' wave operator
\begin{equation}
\square =\delta \mathrm{d}=*\mathrm{d} *\mathrm{d}
\end{equation}
which is just the same on degree $0$ and the same in degree $1$ if we work in a gauge where $\delta =0$. In the rest of this section, we are going to describe some features of the electromagnetic theory arising in this noncommutative context. The electromagnetic theory is the analysis of solutions $A \in \Omega^1(\R^3_\lambda)$ of the equation $\square A=J$ where $J$ is a $1$-form which can be interpreted as a "physical" source. We demonstrate the theory on two natural choices of sources namely an electrostatic and a magnetic one. We start with spin $0$ and we limit ourselves to algebraic plus plane wave solutions.

\section{Spin $0$ modes}
The waves operator on $\Omega^0 (\R^3_\lambda)=\R^3_\lambda$ is comuted from the definitions above as
\begin{equation}
\square =*\mathrm{d}*\mathrm{d}=(\partial^a)^2 
\end{equation}
where the partials are defined by (\ref{partial}). The algebraic massless modes $\ker \square$ are given by \\
\begin{itemize}
\item[$\bullet$] Polynomials of degree one: $f(x)=\alpha +\beta_a x_a$\\
\item[$\bullet$] Linear combinations of polynomials of the type $f(x)=(x_a^2-x_b^2)$\\
\item[$\bullet$] Linear combinations of quadratic monomials of the type, $f(x)=\alpha_{ab} x_a x_b$, with $a \neq b$.\\
\item[$\bullet$] The three particular combinations $f(x)=(2+\delta_{a,1} 10)\lambda x_a^2 -x_1^2 x_a^2$\\
\item[$\bullet$] The three particular combinations $f(x)=(2+\delta_{a,1} 4)\lambda x_a^2 +x_1 x_a^2$\\
\end{itemize}
General eigenfunctions of $\square$ in degree $0$ are the plane waves; the expression for their derivatives can be seen in (\ref{derivativeplanewave}). Hence
\begin{equation}
\square e^{ik.x}= -|k|^2. e^{-2i\lambda k_1} e^{ik.x}
\end{equation} 
It is easy to see that this eigenvalue goes in the limit $\lambda \to 0$ to the usual eigenvalue of the Laplacian in three dimensional commutative space acting on plane waves.

\section{Spin $1$ electromagnetic modes}
On $\Omega^1 (\R^3_\lambda)$, the Maxwell operator $\square_1 =*\mathrm{d} *\mathrm{d}$ can likewise be computed explicitly. If we
writes $A=(\mathrm{d}x_a)A^a$ for functions $A_\mu$, then 
\begin{equation}
F=\mathrm{d}A=\mathrm{d}x_a \wedge \mathrm{d}x_b \partial^b A^a
\end{equation}
If we break this up into  magnetic parts in the usual way then
\begin{equation}
B_a =\epsilon_{abc} \partial^b A^c
\end{equation}
These computations have just the same form as for usual spacetime. The algebraic zero modes $\ker \square_1$ are given by
\begin{itemize}
\item Forms of the type $A=\beta_{ab} (\mathrm{d}x_a)x_b$, wtih $a \ne b$ and curvature 
\begin{equation}
F=\beta_{ab} \mathrm{d}x_a \wedge \mathrm{d}x_b
\end{equation}
\item Forms of the type $A=\gamma x_1 x_a^2$ with curvature
\begin{equation}
F=\gamma \mathrm{d}x_a \wedge \mathrm{d}x_1 x_a^2 
\end{equation}
\item Forms of the type $A=\delta x_1^2 x_a^2$ with curvature
\begin{equation}
F=\mathrm{d}x_a \wedge \mathrm{d}x_1 2 \delta x_1 x_a^2
\end{equation}
\end{itemize}

\section{Magnetic solution}
Here we take a uniform electric current density along a direction vector $k \in \R^3$, i.e. $J=k.\mathrm{d}x=\sum_a k^a \mathrm{d}x_a$. In this case, the gauge potential can be written as
\begin{equation}
A=\frac{1}{4}\left\{ (\sum_{a=1}^3 k_a \mathrm{d}x_a)(C+x_1x_2+x_2 x_3+x_1 x_3) +\sum_{a=1}^3k_a \mathrm{d}x_a x_a^2 \right\}
\end{equation}
The fiels strength is 
\begin{eqnarray}
\nonumber F=\mathrm{d}A&=&\frac{1}{4} \mathrm{d}x_1 \wedge \mathrm{d}x_2 (2k_1 x_2 +k_1 x_1 +k_1 x_3 -2k_2 x_1 -k_2 x_2 -k_2 x_3)\\
\nonumber &+&\frac{1}{4} \mathrm{d}x_1 \wedge \mathrm{d}x_3 (2k_1 x_3 +k_1 x_1 +k_1 x_2 -2k_3 x_1 -k_3 x_3 -k_3 x_2)\\
&+&\frac{1}{4} \mathrm{d}x_2 \wedge \mathrm{d}x_3 (2k_2 x_3 +k_2 x_2 +k_2 x_1 -2k_3 x_2 -k_3 x_3 -k_3 x_1)
\end{eqnarray}
If we decompose the curvature in the usual way then this is an magnetic field in a direction $k \times x$ (the vector cross product). This is a 'confining' (in the sense of increasing with normal distance) version of the field due to a current in direction $k$.
\paragraph{}
We have considered for the electromagnetic solutions only uniform sources $J$; we can clearly put in a functional dependence for the coefficients of the source to similarly obtain other solutions of magnetic types. Solutions more similar to the usual decaying ones, however, will not be polynomial (one would need the inverse of $\sum_a x_a^2$) and are therefore well outside our present scope; even at a formal level the problem of computing d$(\sum_a x_a^2)^{-1}$ in a closed form appears to be formidable. On the other hand these matters could probably be adressed by completing to $\C^*$-algebras and using the functional calculus for such algebras.

\newpage

\section{Discussion}
We choose the representation (\ref{representation}) and $\Lambda =e_3$
because it was convenient when we compute the $1$-form $\mathrm{d}x_1, \mathrm{d}x_2, \mathrm{d}x_3$. None of them was zero which makes only $H^0=\C .1$. It's a good starting point for our model.

\end{document}